\documentclass[aps,twocolumn,floats,preprintnumbers,showpacs,prl]{revtex4}
\usepackage{amsmath,amssymb,epsfig,color,bbold}
\usepackage{feynmf}
\usepackage{feynmf}
\usepackage{psfrag}
\usepackage{graphicx}
\usepackage{bm}

\newcommand{\be}{\begin{equation}}  
\newcommand{\ee}{\end{equation}} 
  
\newcommand{\Dslash}{D\!\!\!\!/\,\,}
\newcommand{\nl}{\nonumber \\ }

\newcommand{\order}{{\cal O}}
\long\def\symbolfootnote[#1]#2{\begingroup
\def\thefootnote{\fnsymbol{footnote}}\footnote[#1]{#2}\endgroup}

\allowdisplaybreaks

\begin{document}

\fmfcmd{
vardef middir(expr p,ang) = dir(angle direction length(p)/2 of p + ang) enddef;
style_def arrow_left expr p = shrink(.7); cfill(arrow p shifted(4thick*middir(p,90))); endshrink enddef;
style_def arrow_left_more expr p = shrink(.7); cfill(arrow p shifted(6thick*middir(p,90))); endshrink enddef;
style_def arrow_right expr p = shrink(.7); cfill(arrow p shifted(4thick*middir(p,-90))); endshrink enddef;}

\preprint{EFI Preprint 13-23}

%\title{Effective field theory for heavy WIMPs scattering with nucleons}
\title{WIMP-nucleon scattering with heavy WIMP effective theory}

\author{Richard J. Hill}
\email{richardhill@uchicago.edu}
\affiliation{Enrico Fermi Institute and Department of Physics, The University of Chicago, Chicago, Illinois, 60637, USA}

\author{Mikhail P. Solon}
\email{mpsolon@uchicago.edu}
\affiliation{Enrico Fermi Institute and Department of Physics, The University of Chicago, Chicago, Illinois, 60637, USA}

\begin{abstract}
The discovery of a Standard Model-like Higgs boson and the hitherto absence of evidence for other new states may indicate that if WIMPs comprise cosmological dark matter, they are heavy compared to electroweak scale particles, $M \gg m_{W^\pm}, m_{Z^0}$. In this limit, the absolute cross section for a WIMP of given electroweak quantum numbers to scatter from  a nucleon becomes computable in terms of Standard Model parameters. We develop effective field theory techniques to analyze the heavy WIMP limit of WIMP-nucleon scattering, and present the first complete calculation of the leading spin-independent cross section in Standard Model extensions consisting of one or two electroweak $SU(2)_W \times U(1)_Y$ multiplets. The impact on scattering cross sections of the choice of WIMP quantum numbers and an extended Higgs sector is investigated.
\end{abstract}
\maketitle

\noindent 
{\bf Introduction.} 
Cosmological evidence for dark matter (DM) consistent with Weakly Interacting Massive Particles (WIMPs) motivates laboratory searches for such particles interacting with nuclear targets. Search strategies and detection potential are highly dependent on the WIMP's properties including its spin, its mass, $M$, and its Standard Model (SM) gauge quantum numbers. In the absence of other signals to guide the search for physics beyond the SM, it is important to identify plausible cross section targets to guide and interpret next generation searches.

The discovery of a SM-like Higgs boson~\cite{Aad:2012tfa} and the hitherto absence of evidence for other new states may suggest that a WIMP, if it exists, is heavy compared  to electroweak scale particles ($M \gg m_{W^\pm}, m_{Z^0}$). In this limit, heavy-particle methods provide theoretical control without assuming a particular ultraviolet completion, allowing us to predict the absolute cross section for a WIMP of given electroweak quantum numbers to scatter from nucleons in terms of SM parameters. This universality is similar to that underlying the predictions of heavy-quark spin symmetry ($m_b\gg \Lambda_{\rm QCD}$) or nonrelativistic atomic spectra ($m_e \gg 1/a_\infty$).   

The SM exhibits a surprising transparency of nucleons to WIMP scattering, due to a cancellation between spin-0 and spin-2 amplitude contributions~\cite{Hisano:2011cs,Hill:2011be}. Robust cross section predictions demand a complete treatment of both perturbative and hadronic uncertainties, including resummation of large logarithms in perturbative QCD (pQCD). In the heavy-particle limit, there is also an intricate interplay between mass-suppressed mixed-state contributions and loop-suppressed pure-state contributions. To analyze these phenomena, we construct the effective field theory (EFT) for heavy WIMPs interacting with SM Higgs and electroweak gauge fields. For the SM extensions under consideration, we present the first computation of the leading $1/M^0$ cross section including matching at leading order in perturbation theory onto the complete basis of operators at the electroweak scale. We summarize here several phenomenological results of this analysis, and present details in companion papers~\cite{Hill:2014yka, part2}. 

\vspace{1mm}
\noindent
{\bf Heavy WIMP effective theory.}
A large class of models, e.g., neutralinos of supersymmetric SM extensions~\cite{Jungman:1995df}, have a WIMP as the lightest state of a new sector. In this situation, the SM is extended at low energies by one or a few particles transforming under definite representations of SM gauge groups. While our analysis is not wedded to supersymmetry (SUSY), SUSY is one of the most-studied SM extensions, and we adopt doublet (``higgsino") and triplet (``wino'') gauge representations as illustrations of ``pure states''. We also consider singlet-doublet (``bino-higgsino") and triplet-doublet (``wino-higgsino") combinations as examples of ``mixed states".

If particles of the new sector are heavy compared to SM particles ($M \gg m_{W}$), we may integrate out the mass scale $M$ using heavy particle EFT. At leading order in the $1/M$ expansion, the heavy-particle lagrangian with time-like reference vector $v^\mu$ is
\begin{equation}\label{eq:heavy}   
{\cal L} = \bar{h}_v \left[
iv\cdot D - \delta m -f(H) \right] h_v  + \order(1/M) \,,  
\end{equation}
where $h_v$ is a heavy-particle field transforming in a representation of electroweak $SU(2)_W$ and $U(1)_Y$, with respective coupling constants $g_2$ and $g_1$. The matrix $f(H)$ describes linear coupling to the Higgs field, and the residual mass matrix $\delta m$ accounts for non-degenerate heavy-particle states.

For extensions with one electroweak multiplet (pure states), the above lagrangian is completely specified by electroweak quantum numbers since gauge-invariance implies $f(H)=0$, and $\delta m$ can be chosen to vanish for degenerate heavy-particle states. In particular, the first term in (\ref{eq:heavy}) does not depend on the WIMP mass, spin or other properties beyond the choice of gauge quantum numbers. Model dependence is systematically encoded in operator coefficients representing $1/M$ corrections. For extensions with two electroweak multiplets (mixed states), $f(H)$ and $\delta m$ are non-vanishing and depend on $\Delta$, the mass splitting of the multiplets, and $\kappa$, their coupling strength mediated by the Higgs field.

\vspace{1mm}
\noindent
{\bf Weak-scale matching.}
Interactions of the lightest, electrically neutral, self-conjugate WIMP, $\chi_v$, with quarks and gluons, relevant for spin-independent (SI), low-velocity scattering with a nucleon, are given at energies $E \ll m_W$ by the EFT
\begin{multline}\label{eq:quarkglue} {\cal L}_{\chi_v,{\rm SM}} =
{\bar{\chi}_v \chi_v \over m_W^3}   
%\bigg\{  
\sum_{S}  \bigg[
\sum_{q} c_{q}^{(S)} O_{q}^{(S)}  
% + c_{q}^{(2)} v_\mu v_\nu O_{q}^{(2)\mu\nu}  \bigg] 
%\\
+ c_{g}^{(S)} O_{g}^{(S)} \bigg]+ \dots  \,,
%+ c_{g}^{(2)} v_\mu v_\nu O_{g}^{(2)\mu\nu}  \bigg\} + \dots  \,,
\end{multline} 
where $q=u,d,s,c,b$ is an active quark flavor and we have chosen QCD quark and gluon operators of definite spin, $S=0,2$:
$O^{(0)}_{q}= m_q \bar{q} q$, 
$O^{(0)}_g = (G^A_{\mu\nu})^2$, 
$O^{(2)\mu\nu}_{q} = \frac12 \bar{q}\left( \gamma^{\{\mu} iD_-^{\nu\}}  - g^{\mu\nu} i\Dslash_-/4 \right) q$, 
and $O^{(2)\mu\nu}_g= -G^{A \mu\lambda} G^{A \nu}_{\phantom{A \nu} \lambda} + g^{\mu\nu}(G^A_{\alpha\beta})^2/4$.
Here $D_-^\mu \equiv \overrightarrow{D}^\mu - \overleftarrow{D}^\mu$,
and $A^{\{\mu}B^{\nu\}} \equiv (A^\mu B^\nu + A^\nu B^\mu)/2$ denotes
symmetrization. The ellipsis in Eq.~\ref{eq:quarkglue} denotes higher-dimension operators suppressed by powers of $1/m_W$.

We match EFTs (\ref{eq:heavy}) and (\ref{eq:quarkglue}) at reference scale $\mu_t \sim m_W \sim m_t$ by integrating out weak scale particles $W^\pm$, $Z^0$, $h^0$ and $t$. In the heavy WIMP limit, matching coefficients, $c_i$, of (\ref{eq:quarkglue}) may be expanded as 
\be
c_i = c_{i,0} + c_{i,1} \frac{m_W}{M} + \dots \, .
\ee
We compute the complete set of twelve matching coefficients $c_{i,0}$ at leading order in perturbation theory. 

Weak-scale matching for mixed states requires renormalization of the Higgs-WIMP vertex for a consistent evaluation of loop-level amplitudes, and a generalized basis of heavy-particle loop integrals to account for non-vanishing residual masses. Details of the matching computation can be found in~\cite{Hill:2014yka}.

\begin{figure}[t] 
\centering
\hspace*{2mm}\includegraphics[scale=0.6]{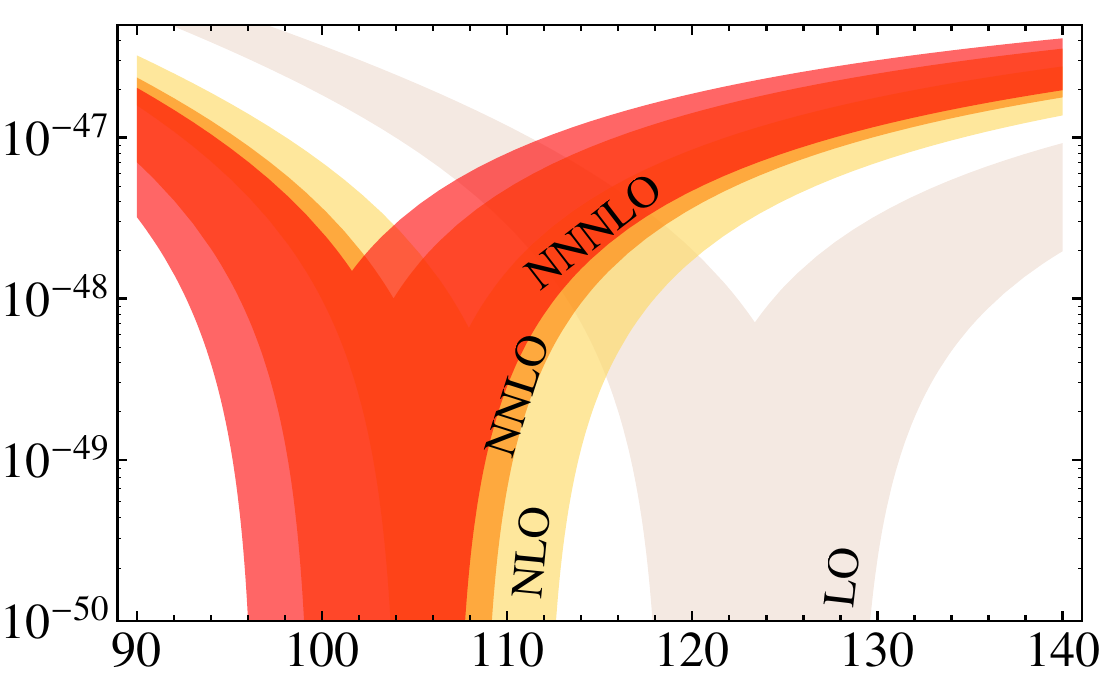}
\centering
\begin{minipage}{0cm}
\vspace*{0.45cm}\hspace*{-6.25cm}$m_h\, ({\rm GeV})$
\end{minipage}
\begin{minipage}{0cm}
\vspace*{-4cm}\hspace*{-14.4cm}\rotatebox{90}{$\sigma_{\rm SI} \, ({\rm cm}^2)$}
\end{minipage}
\caption{\label{fig:winopqcd} SI cross section for
low-velocity scattering on the proton as a function of $m_h$, for the pure-triplet case. Labels refer to inclusion of LO, NLO, NNLO and NNNLO corrections in the RG running from $\mu_c$ to $\mu_0$ and in the spin-0 gluon matrix element. Bands represent $1\sigma$ uncertainty from neglected higher order pQCD corrections.} 
\end{figure}

\vspace{1mm}
\noindent
{\bf QCD analysis.}
Having encoded physics of the heavy WIMP sector in matching coefficients of (\ref{eq:quarkglue}), the remaining analysis is independent of the $M\gg m_W$ assumption, and consists of renormalization group (RG) running to a low scale $\mu_0 < m_c$, matching at heavy quark thresholds, and evaluating hadronic matrix elements. This module is systematically improvable in subleading corrections and is applicable to generic direct detection calculations. An extension of the operator basis would allow robust connections between contact interactions constrained at colliders and low-energy observables of direct detection~\cite{Beltran:2010ww}. RG evolution accounts for perturbative corrections involving large logarithms, e.g., $\alpha_s(\mu_0) \log m_t/\mu_0$. Fig.~\ref{fig:winopqcd} illustrates the impact of higher order pQCD corrections. We collect in Refs.~\cite{Hill:2011be,part2} the details of mapping high-scale matching coefficients onto the low-energy theory where hadronic matrix elements are evaluated
\footnote{
We use updated lattice results for the pion-nucleon sigma term, $\Sigma_{\pi N} = 39 {}^{+18}_{-8} \,{\rm MeV}$~\cite{Durr:2011mp}, and the strange scalar matrix element, $\Sigma_s = 40 \pm 20 \, {\rm MeV}$~\cite{Junnarkar:2013ac}. For $\Sigma_s$ we assume a $50 \%$ uncertainty (cf. the estimated $25 \%$ in \cite{Junnarkar:2013ac}). The up and down scalar matrix elements can be written in terms of $\Sigma_{\pi N}$, $(m_d - m_u) \langle N| (\bar{u} u - \bar{d} d) |N \rangle = 2(2)~{\rm MeV}$~\cite{Gasser:1982ap}, and  $m_u/m_d = 0.49(13)$~\cite{Beringer:1900zz}.}. Cross sections for scattering on the neutron and proton are numerically similar; we present results for the latter.

\vspace{1mm}
\noindent
{\bf Pure-state cross sections.}
Consider the situation where the SM is extended by a single electroweak multiplet. For definiteness let us take the cases of a Majorana $SU(2)_W$ triplet of $Y=0$, and a Dirac $SU(2)_W$ doublet of $Y= \frac12$. For the doublet we assume that higher-dimension operators cause the mass eigenstates after electroweak symmetry breaking (EWSB) to be self-conjugate combinations $D_1$ and $D_2$, thus forbidding a tree-level $\bar{\chi}_v \chi_v Z^0$ coupling, and moreover that inelastic scattering is suppressed.

\begin{figure}[t]
\centering
\hspace*{2mm}\includegraphics[scale=0.6]{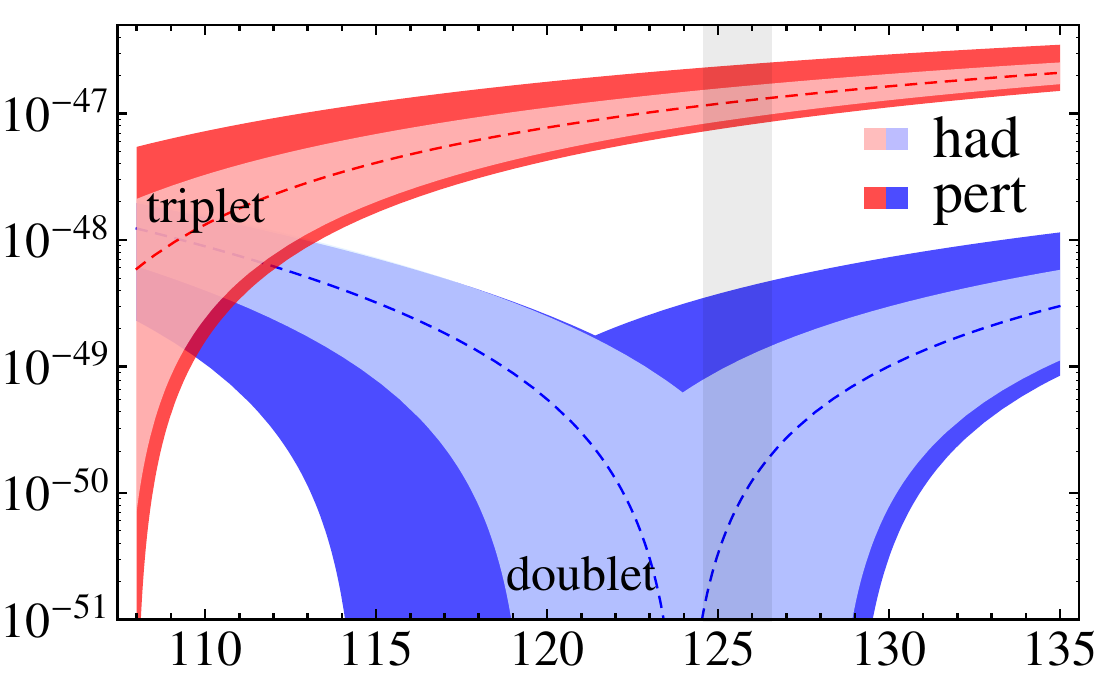}
\centering
\begin{minipage}{0cm}
\vspace*{0.45cm}\hspace*{-6.25cm}$m_h\, ({\rm GeV})$
\end{minipage}
\begin{minipage}{0cm}
\vspace*{-4cm}\hspace*{-14.3cm}\rotatebox{90}{$\sigma_{\rm SI} \, ({\rm cm}^2)$}
\end{minipage}
\caption{\label{fig:purexs}SI cross sections for low-velocity scattering on the proton as a function of $m_h$, for the pure cases indicated.
Here and in the plots below, dark (light) bands represent $1 \sigma$ uncertainty from pQCD (hadronic inputs). The vertical band indicates the physical value of $m_h$.}
\end{figure}

Upon performing weak-scale matching \cite{Hill:2014yka} and mapping to a low-energy theory for evaluation of matrix elements~\cite{part2}, we obtain parameter-free cross section predictions as illustrated in Fig.~\ref{fig:purexs}. The triplet cross section is 
\be\label{eq:triplet} 
\sigma_{\rm{SI}}^T = 1.3 {}^{+1.2}_{-0.5} {}^{+0.4}_{-0.3} 
\times 10^{-47}\,{\rm cm}^2, 
\ee
where the first (second) error represents $1 \sigma$ uncertainty from pQCD (hadronic inputs). Subleading corrections in ratios $m_b/m_W$ and $\Lambda_{\rm QCD}/m_c$ are expected to be within this error budget. Stronger cancellation between spin-0 and spin-2 amplitudes in the doublet case implies a smaller cross section,
\be
\label{eq:doublet} 
\sigma_{\rm{SI}}^D \lesssim 10^{-48}\,{\rm cm}^2 \quad (95\%\,{\rm C.L.}) \,.  
\ee

\begin{figure}[t]
\centering
\hspace*{2mm}\includegraphics[scale=0.6]{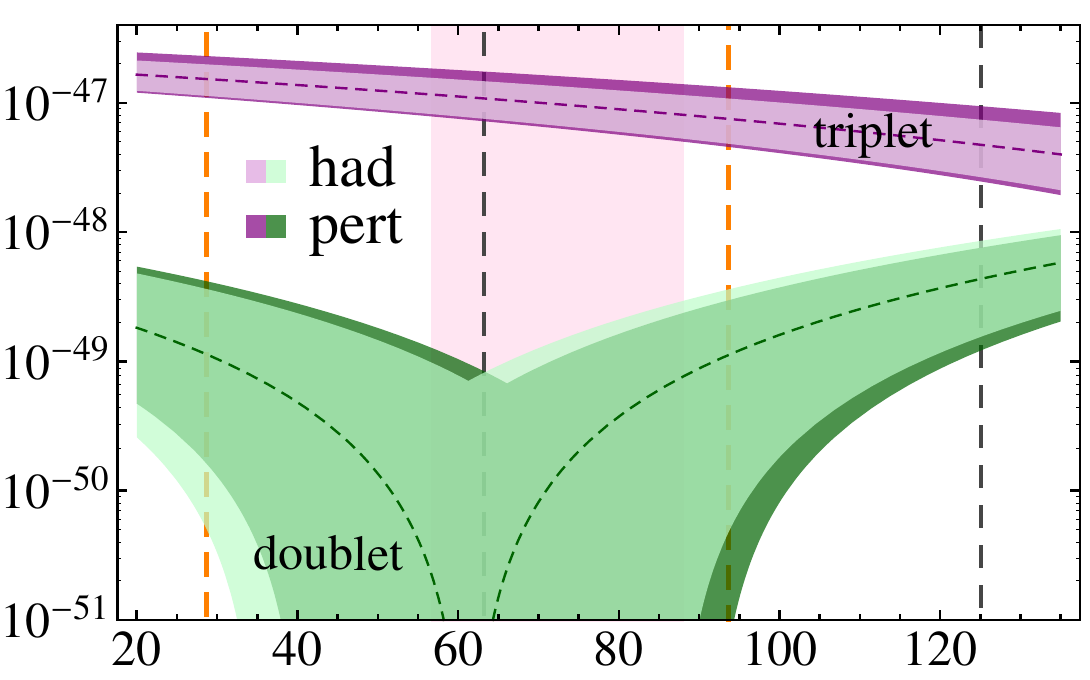}
\centering
\begin{minipage}{0cm}
\vspace*{0.45cm}\hspace*{-6.25cm}$\langle N |m_c \bar{c} c | N \rangle \, ({\rm MeV})$
\end{minipage}
\begin{minipage}{0cm}
\vspace*{-4cm}\hspace*{-14.2cm}\rotatebox{90}{$\sigma_{\rm SI} \, ({\rm cm}^2)$}
\end{minipage}
\caption{\label{fig:charm} SI cross sections for low-velocity scattering on the proton, evaluated in the $n_f=4$ flavor theory as a function of the charm scalar matrix element, for the pure cases indicated.
The pink region corresponds to charm content estimated from pQCD \cite{Junnarkar:2013ac}. The region between orange (black) dashed lines correspond to direct lattice determinations in \cite{Freeman:2012ry} (\cite{Gong:2013vja}).}
\end{figure}

We may also evaluate matrix elements in the $n_f=4$ flavor theory. Figure~\ref{fig:charm} shows the results as a function of the charm scalar matrix element. Cancellation for the doublet is strongest near matrix element values estimated from pQCD. Direct determination of this matrix element could make the difference between a prediction and an upper bound for this (albeit small) cross section.

Previous computations of WIMP-nucleon scattering have focused on a different mass regime where other degrees of freedom are relevant~\cite{Drees:1992rr}, or have neglected the contribution $c_g^{(2)}$ from spin-2 gluon operators~\cite{Hisano:2011cs}. For pure states, this would lead to an $\order(20\%)$ shift in the spin-2 amplitude \footnote{For comparison, neglecting the spin-2 quark contribution from $(b, c, s, d, u)$ shifts the spin-2 amplitude by $\order(1\%, 10\%, 10\%, 30\%, 50\%)$.}, with an underestimation of the perturbative uncertainty by $\order(70\%)$. Due to amplitude cancellations, the resulting effect on the cross sections in Fig.~\ref{fig:purexs} ranges from a factor of a few to an order of magnitude.

\vspace{1mm}
\noindent
{\bf Mixed-state cross sections.}
Mixing with an additional heavy electroweak multiplet (of mass $M^\prime$) can allow for tree-level Higgs exchange, but with coupling that may be suppressed by the mass splitting $\Delta \equiv (M^\prime - M)/2$. We systematically analyze the resulting interplay of mass-suppressed and loop-suppressed contributions through an EFT analysis in the regime $m_W, |\Delta| \ll M,M^\prime$.

Consider a mixture of Majorana $SU(2)_W$ singlet of $Y=0$ and Dirac $SU(2)_W$ doublet of $Y=\frac12$, with respective masses $M_S$ and $M_D$. The heavy-particle lagrangian is given by (\ref{eq:heavy}), where $h_v = (h_S, h_{D_1}, h_{D_2})$ is a quintuplet of self-conjugate fields. The gauge couplings are given in terms of Pauli matrices $\tau^a$,
\begin{align}  \label{eq:bhgauge} 
T^a &= 
\left(\begin{array}{ccc} 0 & \cdot & \cdot \\ 
\cdot &{ \tau^a \over 4} & \, {-i\tau^a \over 4}  \\ 
\cdot & {i\tau^a \over 4} & {\tau^a \over 4}
\end{array} \right) - \text{c.c.}\, , \ \
Y = 
\left(\begin{array}{ccc} 0 & \cdot & \cdot \\ 
\cdot & \mathbb{0}_2 & \, {-i \mathbb{1}_2 \over 2} \\
\cdot & {i \mathbb{1}_2 \over 2} &  \mathbb{0}_2
\end{array} \right).
\end{align} 
The couplings to the Higgs field and residual mass matrix are respectively given by 
\begin{align} \label{eq:bhhiggs}
f(H) &= 
{g_2 \kappa_1 \over \sqrt{2}}\left(\begin{array}{ccc} 
0 & H^T & \, iH^T \\
H & \mathbb{0}_2 & \mathbb{0}_2 \\
iH & \mathbb{0}_2 & \mathbb{0}_2 
\end{array}
\right) 
+ \Bigg[ \, \begin{array}{c} iH \to  H \\ \kappa_1 \to \kappa_2 \end{array} \, \Bigg]
+ \text{h.c. }\, , \nl
\delta m &=\text{diag} (M_S,M_D \mathbb{1}_4) - M_{\rm ref}  \mathbb{1}_5 \, ,
\end{align}
where $M_{\rm ref}$ is a reference mass that may be conveniently chosen. Upon accounting for masses induced by EWSB, we may present the lagrangian in terms of mass eigenstate fields and derive the complete set of heavy-particle Feynman rules; e.g., the Higgs-WIMP vertex is given by $i g_2 \kappa^2 /\sqrt{\kappa^2 + (\Delta/2m_W)^2} \, \bar{\chi}_v \chi_v h^0$ with
$\kappa \equiv \sqrt{\kappa_1^2 + \kappa_2^2}$ and $\Delta \equiv (M_S-M_D)/2$. We may also consider a mixture of Majorana $SU(2)_W$ triplet of $Y=0$ and Dirac $SU(2)_W$ doublet of $Y=\frac12$. Explicit details for the construction of the EFT for these heavy admixtures can be found in~\cite{Hill:2014yka}.

\begin{figure}[t]
\centering
\hspace*{2mm}\includegraphics[scale=0.6]{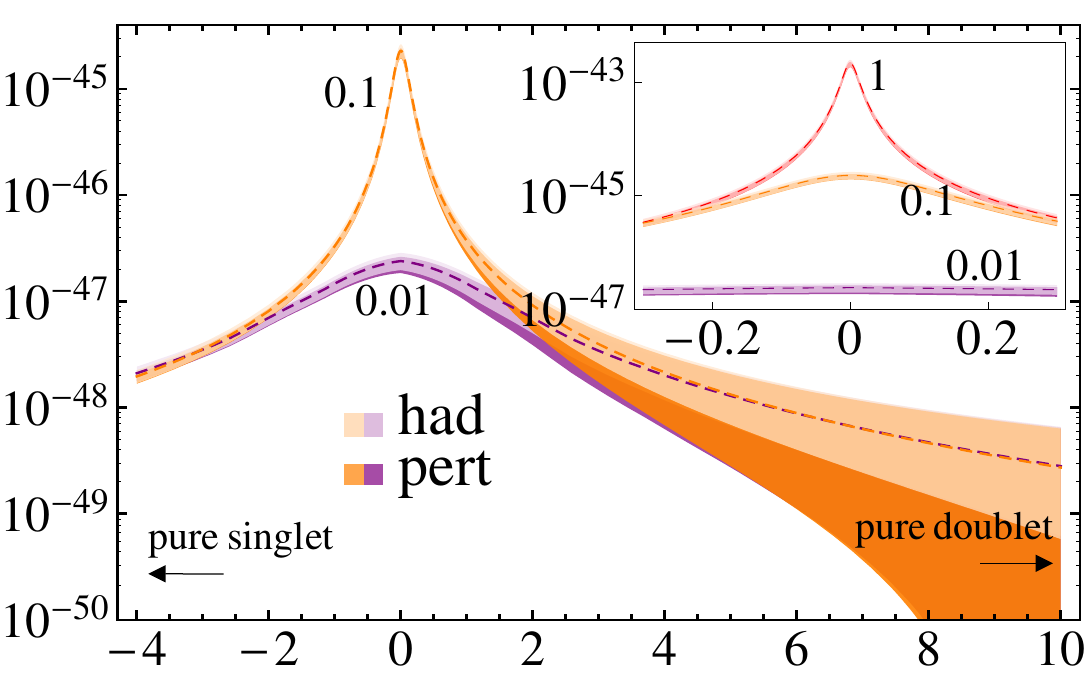}
\hspace*{5.1mm}\includegraphics[scale=0.6]{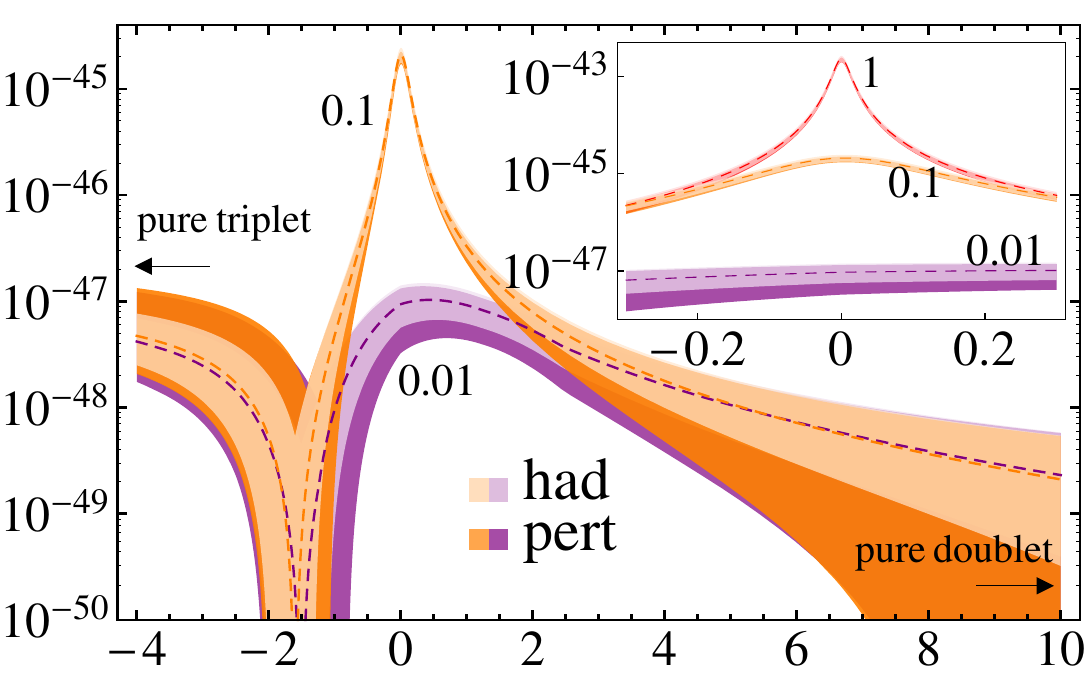}
\centering
\begin{minipage}{0cm}
\vspace*{0.45cm}\hspace*{-6.25cm}$\Delta / [ (4\pi \kappa)^2 m_W ]$
\end{minipage}
\begin{minipage}{0cm}
\vspace*{-12.35cm}\hspace*{-14.3cm}\rotatebox{90}{$\sigma_{\rm SI} \, ({\rm cm}^2)$}
\end{minipage}
\begin{minipage}{0cm}
\vspace*{-4.1cm}\hspace*{-14.5cm}\rotatebox{90}{$\sigma_{\rm SI} \, ({\rm cm}^2)$}
\end{minipage}
\caption{\label{fig:mixed} SI cross sections for low-velocity scattering on the proton for the singlet-doublet and doublet-triplet admixtures, as a function of the mass splitting between pure-state constituents, $\Delta / [ (4\pi \kappa)^2 m_W ]$ (in conveniently chosen units such that interesting features of the curves with different $\kappa$ may be displayed on the same scale). We indicate pure case limits and label each curve with the $\kappa$ value used. Inset plots use the same units.}
\end{figure}

Upon performing weak-scale matching \cite{Hill:2014yka} and mapping to a low-energy theory for evaluation of matrix elements~\cite{part2}, we obtain the results pictured in Fig.~\ref{fig:mixed}. For weakly coupled WIMPs, we consider $\kappa \lesssim 1$. The presence of a scale separation $M, M^\prime \gg m_W$,  implies that the partner state contributes at leading order when $|\Delta| \lesssim m_W$, or more precisely  $|\Delta| \lesssim m_W (4\pi \kappa)^2 $. Within this regime, the purely spin-0 contributions from tree-level Higgs exchange can dominate (cf.~\cite{Cheung:2012qy}). However, when $m_W/\Delta$ suppression is significant, loop-induced contributions become relevant, and the opposite signs of spin-0 and spin-2 amplitudes lead to cancellations in the $\kappa$-$\Delta$ plane. 
In the decoupling limit of SUSY, $\kappa$ depends on $t_\beta$ and the sign of $\mu$, taking values $\kappa \leq \frac12 \tan \theta_W $ ($\kappa \leq \frac12$) for a bino-higgsino (wino-higgsino) mixture. 

\vspace{1mm}
\noindent
{\bf Extended gauge and Higgs sectors.}
A simple dimensional estimate of the pure-state cross section yields $\sigma_{\rm SI} \sim \alpha_2^4 m_N^4/m_W^6  \sim 10^{-45} {\rm \, cm}^2$ \footnote{Results consistent with this estimate were obtained in previous works missing the cancellation~\cite{Cirelli:2005uq, Essig:2007az}.}. However, destructive interference between spin-0 and spin-2 amplitudes leads to anomalously small cross sections. The degree of cancellation depends on SM parameters such as $m_h$ in Fig.~\ref{fig:purexs}, and on the choice of WIMP quantum numbers. Extending our computation to pure states of arbitrary isospin, $J$, and hypercharge, $Y$, the resulting cross section is minimum for $(J,Y) = (\frac12, \frac12)$ corresponding to the doublet, and increases for larger $J$ at fixed $Y$; e.g., the result for $Y=0$ and integer $J$ is $\sigma_{\rm SI}^{(J,0)} = [J(J+1)/2]^2 \sigma_{\rm SI}^T$.

\begin{figure}[t]
\centering
\hspace*{2mm}\includegraphics[scale=0.433]{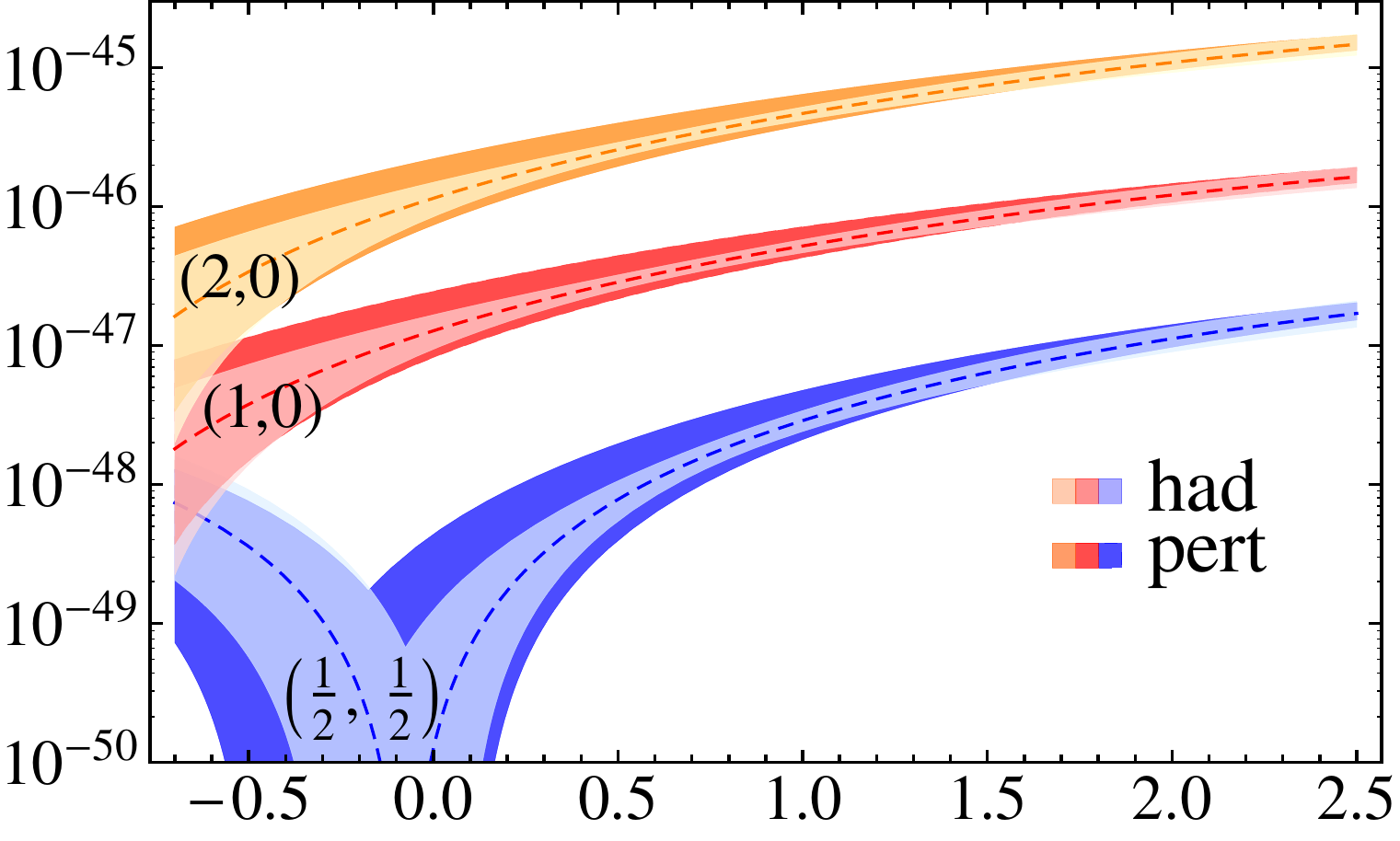}
\centering
\begin{minipage}{0cm}
\vspace*{0.4cm}\hspace*{-6.25cm}$\eta$
\end{minipage}
\begin{minipage}{0cm}
\vspace*{-4cm}\hspace*{-14.2cm}\rotatebox{90}{$\sigma_{\rm SI} \, ({\rm cm}^2)$}
\end{minipage}
\begin{minipage}{0cm}
\vspace*{-1.45cm}\hspace*{-4.9cm}{$\tan \beta = 5\, , \, m_H = 500 \, {\rm GeV}$}
\end{minipage}
\caption{\label{fig:eta} SI cross sections for low-velocity scattering on the proton as a function of $\eta \equiv t_\beta \cos (\beta - \alpha)$, for pure states with quantum numbers $(J,Y)$. The values $|\eta|,|\eta-2| \lesssim 0.5$ are phenomenologically allowed~\cite{Eberhardt:2013uba}. Cross sections assuming only a SM-like Higgs are at $\eta=0$. 
}
\end{figure}

Additional structure in the Higgs sector may also have impact. We illustrate this with a second CP-even Higgs of mass $m_H > m_h = 126 \, {\rm GeV}$, arising in the context of the type-II two-Higgs-doublet model. 
Upon including the contributions of both Higgs bosons, we obtain pure-state cross sections in terms of $m_H$, $t_\beta \equiv \tan \beta$ and $\eta \equiv t_\beta \cos(\beta -\alpha)$ (following the parameterization in Ref.~\cite{Carena:2013ooa} for departures from the ``alignment limit"). For $t_\beta \gg 1$ and $| \eta | \leq \order(1)$, the couplings of the SM-like Higgs to $W^\pm,Z^0,u,c,t$ are given by $1 + \order(1/ t_\beta^2)$, while those to $d,s,b$ are given by $(1- \eta ) + \order(1/ t_\beta^2)$, measured relative to SM values. 
Existing phenomenological constraints are not sensitive to the sign of the latter, allowing for both $\eta \approx 0,2$ where the magnitude is near the SM value~\cite{Eberhardt:2013uba}. 
Figure~\ref{fig:eta} shows cross section predictions for pure states with quantum numbers $(J,Y)$ indicated, including $(2,0)$, the smallest representation for which WIMP decay by dimension five operators is forbidden by gauge invariance~\cite{Cirelli:2005uq}. 

\vspace{1mm}
\noindent
{\bf Discussion.} 
We constructed the EFT for heavy WIMPs interacting with SM gauge and Higgs bosons, and used it to compute predictions with minimal model dependence for cross sections to be probed in future DM search experiments. We presented absolute predictions for WIMPs transforming under irreducible representations of $SU(2)_W \times U(1)_Y$ (Fig.~\ref{fig:purexs}), and considered the impact of additional WIMPs (Fig.~\ref{fig:mixed}) and of an extended Higgs sector (Fig.~\ref{fig:eta}). We also demonstrated the significance of corrections from pQCD (Fig.~\ref{fig:winopqcd}) and of potential improvements in lattice studies of hadronic matrix elements (Fig.~\ref{fig:charm}).

The formalism for weak scale matching computations and QCD effects in general direct detection scenarios are presented in \cite{Hill:2014yka,part2}. The basis of heavy-particle loop integrals arising in heavy WIMP-nucleon scattering can also be applied to low-energy lepton-nucleon scattering \cite{Hill:2012rh}. It is interesting to investigate the impact of nuclear effects on the cancellation between spin-0 and spin-2 amplitude contributions \cite{Prezeau:2003sv}. Hadronic uncertainties are dominated by the strange scalar matrix element~\cite{Durr:2011mp,Junnarkar:2013ac,Freeman:2012ry,Ellis:2008hf}, and within the bounds from current lattice data~\cite{Freeman:2012ry,Gong:2013vja}, a precise determination of the charm scalar matrix element can also have significant impact.

While in general model-dependent, it is interesting to extend the EFT analysis here to include power corrections in specific ultraviolet completions, and incorporate constraints on heavy WIMPs from other observables such as indirect detection~\cite{Cohen:2013ama}.

\vspace{1mm}
\noindent
{\bf Acknowledgements.}
We thank B.~Batell, V.~Cirigliano, R.~Essig, J.~Ruderman, T.M.P.~Tait, C.E.M.~Wagner, and A.~Walker-Loud for insightful discussion. This work was supported by the U.S. DOE under Grant No. DE FG02 90ER40560, in part by the NSF under Grant No. PHYS-1066293, the hospitality of the Aspen Center for Physics, and a Bloomenthal Fellowship.

\end{document}